# Multi-scale Molecular Simulations on Respiratory Complex I

**Ville R. I. Kaila**

Department Chemie, Technische Universität München, Lichtenbergstraße 4, 85747 Garching, Germany. E-mail: ville.kaila@ch.tum.de

**Content**

**Abstract**

Complex I (NADH:ubiquinone oxidoreductase) is a redox-driven proton pump that powers synthesis of adenosine triphosphate (ATP) and active transport in most organisms. This gigantic enzyme reduces quinone (Q) to quinol (QH$_2$) in its hydrophilic domain, and transduces the released free energy into pumping of protons across its membrane domain, up to *ca*. 200 Å away from its active Q-reduction site. Recently resolved molecular structures of complex I from several species have made it possible for the first time to address the energetics and dynamics of the complete complex I using multi-scale methods of computational biochemistry. Here it is described how molecular simulations can provide important mechanistic insights into the function of the remarkable pumping machinery in complex I and stimulate new experiments.

# 1. Introduction to structure and function of complex I

Complex I is one of the largest (0.5-1 MDa) and most intricate respiratory enzymes. It catalyzes electron transfer (eT) from nicotine amide dinucleotide (NADH) to quinone (Q) in its hydrophilic domain along a *ca.* 100 Å wire composed of flavin mononucleotide (FMN) and 8-9 FeS centers.[1-5] This reduces Q to quinol ($QH_2$) in a process that is coupled to pumping of four protons across the membrane domain of complex I,[6-7] up to *ca.* 200 Å away from the site of Q reduction (Figure 1). Despite recently resolved X-ray[8-12] and cryo-EM[13-15] structures from several species, data from labeling-[16-18], crosslinking-,[19] and site-directed mutagenesis studies,[20-26] as well as biophysical experiments,[27,28] the mechanism by which complex I catalyzes this remarkable long-range proton-coupled electron transfer (PCET) process still remains unclear. Elucidating the molecular mechanism of complex I is not only crucial for understanding primary energy transduction in biology, but it is also of great biomedical relevance, since almost half of all known mitochondrial disorders are linked to mutations in complex I.[1,2,29]

The membrane domain of complex I comprises three antiporter-like subunits, NuoN (*E. coli* nomenclature, Nqo14 in *T. thermophilus*, ND5 in human/*Bos taurus*), NuoM (Nqo13/ND4), and NuoL (Nqo12/ND2), each of which contains two pseudo-symmetric trans-membrane (TM) segments, TM4-8 and TM9-13 (Figure 1, *inset*).[9,10,12,14,15] Helices TM7 and TM12 are broken by short loops, and may participate in the pumping process, similar to the structurally related carrier-type transporters, which employ such motifs to transport ions across the membrane.[30] The X-ray structures also show that despite a low sequence similarity, TM2-6 of subunit NuoH (Nqo8/ND1) is structurally similar to the antiporter-like NuoN/M/L subunits.[10] The membrane domain comprises a chain of buried titratable residues (Figure 1), which are central for the proton pumping process.[20-26] The membrane domain is connected by an unusual long transverse HL-helix, originally suggested to provide a piston function that is involved in establishing proton pumping across the membrane.[9] Recent work, however, indicates that the HL-helix may function as a molecular clamp that connects the antiporter-subunits together.[31, but *cf.* also 32, 33]

The proton pumping in complex I is highly efficient, employing up to 97% of the redox potential gap between NADH (-320 mV) and Q (+90 mV in the membranes, but see below).[34] The pumping machinery is fully reversible and electrons can also be extracted from $QH_2$ to drive the reverse eT along the FeS chain to $NAD^+$ by using an external pH gradient across the membrane.[1-5] Such reverse eT conditions are also of physiological relevance, since this increase the production of reactive oxygen species (ROS).[1,2,29] The Q reduction and proton pumping are strongly coupled, and mutations of residues in the terminal NuoL subunit therefore also inhibit the Q-reduction activity,[25,26] as expected based on the principles of microscopic reversibility. Nevertheless, the molecular principles of this coupling remain poorly understood.

Elucidation of molecular structures of complex I in recent years has opened up mechanistic studies of the enzyme for molecular simulations.[32,35-43] These techniques can provide a powerful methodology to study the structure, function, and dynamics of complex biological systems on a wide range of timescales and spatial resolutions. Molecular simulations have played an important role in bioenergetics, providing mechanistic insight into the function of, *e.g.*, cytochrome *c* oxidase,[44-55] photosystem II,[56-62] cytochrome $bc_1$,[63,64] $F_oF_1$-ATPase,[65-68] as well as light-driven ion-pumps.[69,70] Here I describe how methods of computational biochemistry can be used to study the function of complex I, and central information that such simulations may provide. In section 2, the basic theory of the methods, which have been employed in computational studies of complex I, as well as techniques that may provide important input for future work are briefly reviewed, followed by discussion on computational results on complex I in sections 3-8.

## 2. Computational models and methods

### 2.1 Classical Molecular Dynamics (MD) simulations

The goal of classical Molecular Dynamics (MD) simulations is to model molecular interactions to describe the microscopic dynamics of the system of interest. To obtain a computationally efficient evaluation of inter- and intra-molecular interactions, a pre-parameterized force field potential is employed,[71-73]

$$V = \sum_{\text{bonds}} \frac{1}{2} k_b (r - r_0)^2 + \sum_{\text{angles}} \frac{1}{2} k_\theta (\theta - \theta_0)^2 + \sum_{\text{improper}} \frac{1}{2} k_\chi (\chi - \chi_0)^2 + \sum_{\text{dihedrals}} k_\phi [1 + \cos(n\phi - \delta)] + \sum_{i>j} (q_i q_j / 4\pi\varepsilon_0 r_{ij}) + \sum_{i>j} \varepsilon_{ij} [(\sigma_{ij}/r_{ij})^{12} - 2(\sigma_{ij}/r_{ij})^6], \quad (1)$$

where bond-, angle-, dihedral-, and improper- terms describe bonded (covalent) interactions within the molecule, and allow for fluctuation around some equilibrium reference values ($r_0$, $\theta_0$, $\chi_0$, $\phi$). The non-bonded interactions within and between molecules are modeled using a Coulombic electrostatic potential ($q_i$ -point charges; $\varepsilon_0$ vacuum permittivity) and a Lennard-Jones 12-6 term ($\varepsilon_{ij}$ - potential depth; $\sigma_{ij}$ - combined van der Waals radii for atoms $i$ and $j$) to describe dispersive interactions. The functional form of the force field expression allows the molecule to undergo conformational changes, but covalent bonds cannot form or break due to the employed harmonic approximation, as the energy increases parabolically upon displacement of the atoms from their equilibrium values. Biomolecular force fields, such as CHARMM,[71] GROMOS,[72] and AMBER,[73] are parameterized using quantum mechanical calculations and experimental data in order to reproduce, *e.g.*, structural data, diffusion properties, and solvation free energies. Force field parameters are available for common biomolecules, such as amino acids, lipids, nucleic acids, and sugars.[71-73] Complex I, however, comprises several co-factors, including 8-9 iron-sulfur centers, FMN, NADH, and Q that can reside in different redox and protonation states during the catalytic cycle. For these co-factors, standard force field parameterizations are not available, and they thus require parameterization based on quantum chemical calculations (see below). For modeling transient catalytic states of complex I, the cofactors are parameterized in different charge states, which can be used for probing the dynamics of the systems, *e.g.*, prior and after reduction of the co-factors. In addition to so-called type-I force fields (*Eqn*. 1), polarizable models,[74] where the point charge distribution can fluctuate, and reactive force fields,[75,76] where the bonding topology is parameterized to change during the dynamics, have also been developed. Such simulation methods have not yet been employed in studies of complex I, but they may provide important mechanistic insight in the future work. For example, empirical valence bond (EVB) simulations[50,51] have contributed to our understanding of the proton pumping in cytochrome *c* oxidase. In addition to the atomistic MD simulations discussed here, coarse-grained (CG) force fields[77] that model residue interactions on a bead-level instead of considering explicit atomic interactions, are also being developed, and provide access to longer simulation time-scales, *i.e.*, milliseconds rather than microseconds.

In MD simulations, the Newtonian equations of motion (EOM),

$$\mathbf{F} = -\nabla V = m\mathbf{a} = m\frac{d^2 r}{dt^2} \qquad (2)$$

are discretized and integrated numerically. The Verlet algorithm, for example, can be used to calculate the atomic positions in the next time step, $\mathbf{r}(t+\Delta t)$, based on their current, $\mathbf{r}(t)$, and previous positions, $\mathbf{r}(t-\Delta t)$,

$$\mathbf{r}(t+\Delta t) = 2\mathbf{r}(t) + \mathbf{r}(t-\Delta t) + (f(t)/m)\,\Delta t^2 \qquad (3)$$

with forces, $f(t)$, and masses, $m$, of the particles. The forces between atoms are obtained from the gradient of *Eqn.* 1, and initial velocities at $t=0$ are assigned from a Maxwell-Boltzmann velocity distribution, for a given simulation temperature. In order to properly model molecular collisions, the EOMs are integrated using a short time step, $\Delta t$, of 1-2 fs, which captures the fastest bond-vibrations within the system. The time-ordered atomic positions obtained by integrating the EOMs give the MD trajectory, which contains central dynamic information of the system.

The bottleneck in the classical MD simulations are estimations of the long-range electrostatic interactions, since the interaction of each $N$ atom needs to be evaluated with all $N$-1 atoms in the system. As an MD simulation setup of the bacterial complex I comprises *ca.* $10^6$ atoms (see below), this would result in evaluation on the order of $10^{12}$ interactions, which is not possible to achieve even with the fastest computers. The $N^2$-scaling of such computation can, however, be reduced by using, *e.g.*, the Particle Mesh Ewald (PME) algorithm, in which interactions between atoms far in space are calculate in reciprocal (Fourier)-space and added to their direct space contributions, using a fast-Fourier transformation (FFT) algorithm. This procedure lowers the computational scaling to $N\log(N)$-scaling for a system with $N$ atoms. In classical molecular simulations, the temperature ($T$) and pressure ($p$) are also modeled by using thermostats and barostats and are normally set to $T=310$ K and $p=1$ bar in biomolecular modeling. To this end, *Eqns.* 2-3 are modified to the Langevin equations, which also take into account stochastic Brownian forces within the system.

Due to the tremendous increase in computational power in recent years, MD simulations allow today access to microsecond timescales. Special computers, such as the ANTON supercomputer, have pushed this limit even further, allowing simulations to be carried out on the millisecond timescales, at least for small proteins.[78] Information about processes taking place beyond the microseconds timescale can also be obtained from shorter simulations. To this end, non-equilibrium or transient catalytic states and their relaxation can be studied or data is collected from several independent simulations using, *e.g.*, Markov-State models (MSM)[79] that project out molecular motion based on different timescales. Moreover, instead of following direct brute-force dynamics of an individual system, free-energy simulation techniques[80] can be employed to perturb the system to systematically undergo rare conformational changes, followed by reconstruction of the underlying free energy landscape using methods of statistical mechanics (see below). Reaction rates can this way be related to the free energy barriers using transition state theory.

To study the dynamics of complex I, a realistic computational model of the enzyme in its biological surrounding must be built. This starts by modeling all hydrogen atoms that are commonly not resolved in the protein X-rays structures. To this end, protonation states for all titratable residues are assigned by performing, *e.g.*, Poisson-Boltzmann (PB) continuum electrostatic calculations with Monte-Carlo sampling (see below).[81,82] After all atoms have been explicitly modeled, complex I is inserted in a lipid bilayer, and solvated with water molecules and ions. Either single component lipid models, such as 1-palmitoyl-2-oleoyl-sn-glycero-3-phosphocholine (POPC), or multi-component lipids that mimic the composition of the inner mitochondrial membrane, are commonly used in membrane protein simulations.[35-37] The molecular simulation setup of complex I from *Thermus thermophilus* results in a system with *ca.* 840,000 atoms, shown in Figure 1. Although the eukaryotic complex I contains 46 subunits and has a molecular mass of *ca.* 1 MDa, the MD model for the *Bos taurus* enzyme results in only a slightly larger system of *ca.* 1.1 million atoms, since a larger part of the simulation box comprises water molecules. Unfortunately, *ca.* 5% of the residues in the eukaryotic complex I structures still remain unresolved. Although these missing parts can be modeled using protein prediction methods,[83,84] such simulations may have larger uncertainties in comparison to simulations performed based on fully experimentally resolved structures. After the structure of complex I has been inserted in the membrane-water-ion environment, the model is energy minimized and equilibrated followed by production simulations. Up to date, microsecond-timescale MD simulations for several independent trajectories have been reached.[35-37] Simulation of such timescales, requires *ca.* $5 \times 10^8$ integration steps ($\Delta t$=2 fs, *Eqn.* 3) for each of the *ca.* $10^6$ atoms, and therefore access to high-performance supercomputers is necessary.

**2.2 Free-energies and electrostatic Poisson-Boltzmann calculations**

It can be difficult to observe rare-events with high-energy barriers in classical MD simulations. Sampling of such events can, however, be achieved by applying external potentials on a reaction-coordinate of interest that flattens its free-energy landscape. The unbiased free energy profile, *i.e.*, the probability of observing the event of interest without introducing external forces, is obtained from the probability distribution of the reaction coordinate observed in the biased (restrained) simulations by re-weighting the statistics with the employed biasing potential. Commonly used computational free-energy approaches are, *e.g.*, umbrella sampling, metadynamics, and free energy perturbation methods (for further discussion see, *e.g.*, Ref.[80]). Generating converged free energy profiles requires a careful choice of the reaction-coordinate(s) as well as a significant statistical overlap in these coordinates. Application of free energy simulation techniques can thus be very challenging for large systems such as complex I.

The electrostatic free-energy for protonation ($pK_a$) and reduction ($E_m$) processes are particularly relevant for understanding the function of complex I. These properties can also be estimated by solving the (linearized) Poisson-Boltzmann (PB) equation, which relates an electrostatic potential ($\psi$) to a charge distribution ($\rho$),[81,82]

$$\nabla \varepsilon(r) \cdot \nabla \psi(r) - \varepsilon(r) \lambda(r) \left[ \frac{8\pi I q^2}{\varepsilon(r) k_B T} \right] \psi(r) = -4\pi \rho(r), \qquad (4a)$$

where $\varepsilon(r)$ is the dielectric constant, $\lambda$ is a switching function, $I$ is the ionic strength, $k_B T$ is the thermal energy, and $q$ are the point charges in the system. Integration of the obtained potential at the point charges gives the electrostatic free energy,

$$\Delta G = \frac{1}{2}\sum_i q_i \psi_i. \tag{5}$$

To calculate $pK_a$ and/or $E_m$ values, a thermodynamic cycle is employed, where the electrostatic free energies resulting from transfer of the protonated (AH) and deprotonated (A$^-$) (or reduced/oxidized) species from aqueous ($\varepsilon$=80) phase to the protein interior ($\varepsilon$=4-10) are estimated. In addition to this so-called Born desolvation energy, the interaction of AH and A$^-$ is computed with all protein background charges. Inserting AH/A$^-$ into the protein might also result in changes of the $2^N$ possible protonation states of the protein with $N$ titratable residues. Due to the high dimensionality of these possible protonation states, the last effect is often calculated by Monte Carlo sampling. For example, the membrane domain complex I, contains *ca.* 350 titratable residues, giving $2^{350}$ possible protonation states. The electrostatics shifts are commonly calculated relative to an experimentally measured $pK_a$ or $E_m$ values of the model compound, *e.g.*, in water where the values are known to high accuracy.

### 2.3 Quantum Chemical Density Functional Theory (DFT) Models

In order to capture the energetics of a chemical process, *e.g.*, Q reduction to QH$_2$ or proton transfer across the membrane domain, the motion of electrons must be rigorously described based on the Schrödinger equation. Although the exact solution of this equation is not possible for large molecules, density functional theory (DFT) calculations provide an accurate methodology to approximate the many-particle Schrödinger equation, with a good balance between computational cost and accuracy. DFT is in principle exact in the non-relativistic limit, thus representing a rigorous reformulation of the Schrödinger equation, and awarded its developer Walter Kohn a Nobel Prize in 1997. The exact dependence of how the energy is related to the electron density is, however, still unknown, and it was not until the early 1990s when accurate approximations of the so-called exchange-correlation term were developed. One commonly employed density functional is Becke's three-parameter hybrid functional, B3LYP,[85,86] which has become important in computational biochemistry.[38,44,47,56-58,61,62] In this functional, an empirical amount (20%) of Pauli electron-exchange is introduced from the mean-field Hartree-Fock theory. Different density functionals have a benchmarked error of *ca.* 1-5 kcal mol$^{-1}$ in reaction energies, whereas geometries are usually predicted within an accuracy of 0.05 Å (in bond distances) relative to experimental geometries on model compounds.[87] Despite these and other systematic errors (see *e.g.* Ref.[87]), DFT nevertheless remains one of the most powerful and accurate techniques to treat complex biochemical systems at a QM level.

In order to perform DFT calculations, the electrons are modeled by finite basis sets, constructed from a linear-combination of atomic orbitals to give molecular orbitals (LCAO-MO). Basis sets are often composed of a sum of Gaussian functions, and benchmarking studies suggest that basis-set convergence within DFT is reached when the electrons are modeled using a triple set of functions (triple-zeta quality basis set, *e.g.* def2-TZVP or 6-311G**+).[88] The basis sets are used to solve the Kohn-Sham equations in DFT, until self-consistent solutions are obtained. This procedure gives molecular orbitals with optimized weight of each orbital contribution, as well as a total energy for the system that can be compared between different states for geometry-optimizes structures. DFT calculations allow today treatment of systems with *ca.* 100-500 QM atoms and can be used for structure or reaction pathway optimization, first-principles dynamic simulations, or molecular property calculations.

Care must be taken when treating certain properties at DFT level that result from electron correlation effects.[87] DFT is not able to rigorously capture dispersion interactions, but this central interaction is modeled in dispersion-corrected density functionals (DFT-D),[89] by introduction of an empirical $r^{-6}$ dispersion term to the functional, and is commonly used in computational biochemistry. Another challenging problem central for biochemical system, and particularly for the FeS centers in complex I, is the accurate treatment of spin energetics. It is known that, *e.g.*, the B3LYP functional has a tendency to systematically overestimate the stability of high-spin configurations,[90] as the high spin energy becomes favored by increasing the amount of introduced exchange, an empirical parameter within the hybrid density functionals. It can therefore be important to benchmark the performance of different functionals on the property of interest. The spin-energetics of the FeS centers in complex I have a particularly challenging electronic structure, as each of the irons have 4-5 unpaired electrons, which are anti-ferromagnetically coupled together to yield a $S=1/2$ or 0 system. For the tetranuclear FeS centers (N2, N7, N6a, N6b, N5, N4, N3), there are six unique anti-ferromagnetically coupled spin-states that each may have different energies. In order to achieve proper spin states, the broken-symmetry spin-flip approach can provide a particularly useful approximation in estimating spin-coupling parameters within the framework of DFT.[91,92]

In order to study a protein function using DFT methods, a QM cluster model, capturing the central chemical environment can be constructed.[93] In such models, the active system of interest together with central first- and second sphere residue interactions, are cut out from the biological environment. The terminal atoms are saturated with hydrogen atoms, and fixed during structure optimization to mimic strain that arises form the protein environment. Moreover, the electric response of the environment can be modeled using implicit dielectric polarizable medium models with a dielectric constant usually set at 4-20. The total energy of the QM cluster models in different states can be systematically compared, as proper structure optimizations can be performed. When the system size becomes very large, however, it is increasingly difficult to find proper energy minima, leading to uncertainties in the total energy of the system. Such problems may also arise when explicit dynamics of the system is considered.

**2.4 Hybrid quantum mechanics/classical mechanics (QM/MM) models**

In hybrid quantum mechanics/classical mechanics (QM/MM) calculations the QM system of interest is embedded and polarized by a protein surroundings, which is described at the classical force field level.[94] In so-called additive QM/MM schemes, the total energy and forces are calculated from the sum of the QM and MM subsystems and the interactions between the QM and MM parts ($E_{QM/MM} = E^{MM} + E^{QM} + E_{QM-MM}$).[95] There are also subtractive QM/MM schemes, *e.g.*, the ONIOM method ("our own *N*-layered integrated molecular orbital and molecular mechanics"),[96] where the total classical energy is corrected by the energy differences between the MM and QM parts for the central region of interest ($E_{QM/MM} = E_{large}^{MM} - E_{small}^{MM} + E_{small}^{QM}$; where "*large*" refers to the complete system and "*small*" is the central region of interest). In QM/MM calculations, the QM region is defined and separated from its classical surroundings by, *e.g.*, cutting between Cα and Cβ atoms for each residue involved, and introducing link atoms between these regions that chemically saturate the QM system. The link atoms are made invisible for the MM region, and charge distributions at the boundary regions are spread out to prevent over-polarization effects. In QM/MM simulations, the energy and forces for the central QM region are calculated *on-the-fly* using, *e.g.,* DFT, which replaces the pre-parameterized force field expression (*Eqn.* 1) for this region. It can be confusing for a non-expert to distinguish between different QM/MM simulations as the QM treatment may refer to DFT, *ab initio* theory, or semi-empirical methods. For example, the self-consistent charge-tight binding density functional theory (SCC-DFTB) has provided mechanistic insight into the function of bacteriorhodopsin[69] and cytochrome *c* oxidase.[54] In SCC-DFTB, the electron density is replaced by point-charges instead of the electron density as in DFT. Although it is difficult to benchmark the exact computational errors of different QM/MM methods, DFT-based QM/MM simulations normally allow accessing some tens of picoseconds timescale, whereas semi-empirical QM/MM can normally be extended to the 100 ps - 1 ns timescales.

### 3. Dynamics of electron transfer

While NADH is a two electron carrier, the FeS centers in the hydrophilic domain of complex I undergo one-electron oxidoreduction, switching between their reduced ($2Fe^{2+}2Fe^{2+}$ or $Fe^{2+}Fe^{2+}$) and oxidized ($3Fe^{2+}Fe^{3+}$ or $Fe^{3+}Fe^{2+}$) forms.[97] The hydrophilic domain of complex I thus functions as a "*two-to-one*" electron converter, that bifurcates the eT from NADH to Q. The first electron from NADH is transferred by PCET via FMN to the binuclear N1a center. Flavosemiquinone species ($FMN^{-/\bullet}$ or $FMNH^{\bullet}$) have not yet been observed, and the mechanism for this putative hydride ($H^{\bullet/-}$) transfer also remains unclear.[2,5] Experiments, however, show that after reduction of N1a, the second electron rapidly moves from NADH/FMN along the *ca*. 100 Å chain of FeS clusters to the high potential N2 center in *ca*. 20 μs,[27,28] while the resulting $NAD^+$ is kinetically trapped,[98] possibly to avoid leaking the electron from N1a to oxygen in bulk solvent. All FeS clusters except N7, which resides *ca*. 20 Å from the main eT pathway,[27] participate in the transfer process, and the FeS centers range in redox mid-point potentials ($E_m$) from -330 mV to *ca*. -200 mV (N2). Experiments further show that reduction of N2 results in a slower (millisecond) redistribution of the electron from N1a to the other FeS centers.[27]

The rate for this eT process can be estimated from the empirical Moser-Dutton ruler,[99]

$$\log k_{eT} = 13 - (1.2 - 0.8\rho)(R\,[\text{Å}] - 3.6) - 3.1[\text{eV}^{-1}]\frac{(\Delta G - \lambda)^2}{\lambda} \qquad (6)$$

with free energies of the inter-FeS center eT processes ($\Delta G$) derived from electrochemical experiments, *edge-to-edge* distances ($R$) from X-ray structures, and by employing typical reorganization energies (λ=0.5-0.7 eV), and protein packing densities (ρ=0.76). This gives an eT rate in the milliseconds timescale, [2,5,100] which is somewhat slower or comparable to the overall complex I turnover of *ca*. 2 ms. It is, however, challenging to assign the Δ*G*s based on experiments, since the electrostatic couplings between the centers are not directly obtained from the measured redox potentials. To this end, Couch *et al*.[40] and Medvedev *et al*.[39] calculated electrostatic couplings between the FeS centers based on PB electrostatics calculations and experimental constraints from the measured $E_m$ values. These "coupling"-corrected $E_m$ values obtained using different dielectric environments (ε=4-20) upshift some FeS redox potentials, and give eT rates of *ca*. 0.4-4 ms based on a Moser-Dutton treatment.[5] Hayashi *et al*.[41] also addressed whether tunneling pathways could increase the overall transfer rate, by explicitly calculating the electronic overlaps along different pathway.

By performing semi-empirical calculations at the broken-symmetry ZINDO/S level on the FeS centers and their nearby surroundings, they found that many of the pair-wise transfer processes are indeed faster than the overall turnover. However, these calculations suggested that the transfer between N5 → N6a ($10$ s$^{-1}$) and N3 → N1b ($10^3$ s$^{-1}$) form bottlenecks in the process. They further found that internal water molecules, which have not yet been resolved in the experimental structures of complex I, may increase the eT rate by bridging unfavorable gaps between the FeS centers and increase the electronic coupling along the pathways. MD simulations on the microseconds timescale suggest that some water molecules can indeed enter the hydrophilic domain, *e.g.*, the region between N5 and N6a (Figure 2A). The simulations, however, also show that there are relatively large fluctuations in many of the inter-FeS distances of *ca*. 2-3 Å (Figure 2B). In addition to the uncertainties in $\Delta G$s and $\lambda$s, fluctuations in $R$ may also modulate the eT rate by several orders of magnitude, as shown for the rate-limiting eT step between N5 and N6a in Figure 2C. Moreover, although many of the predicted tunneling rates[41] are indeed consistent with the experimental turnover-constraints, a potential error source in these pathway calculations could also arise from the treatment of the highly challenging FeS spin-energetics (see section 2.3). A DFT treatment could offer a more accurate, yet computationally expensive option to probe the spin-energetics relative to semi-empirical methods, whereas development of novel quantum chemical multi-reference methods, such as the density-matrix renormalization group (DMRG) theory,[101] might in future open up an accurate correlated *ab initio* treatment of such challenging electronic structure problems.

Interestingly, de Vries *et al.*[28] recently found that the eT between N5 → N6a (N4Fe[75]H →N4 in *E. coli*) becomes sixfold slower upon reduction of N2, suggesting that complex I might utilize a feedback regulation mechanism to modulate the rate of Q reduction, possibly in order to time it with the pumping cycle. A direct Coulombic-interaction between N2 and N6a, would be expected to tune the $\Delta G$ for this eT by *ca*. 60 mV, decreasing the eT rate by *ca*. 20%, which may account for this effect in part. PB calculations[38] and experiments,[102] suggest that His-169 (*T. thermophilus* numbering, if not otherwise stated) becomes protonated upon reduction of N2. Such effect might change the local electrostatic couplings by conformational changes in surrounding charged residues near the terminal FeS centers, and modulate the overall eT rate. Although the molecular mechanism for this putative regulation process remains unclear, molecular simulations can be used to predict eT-parameters,[103] thus establishing a molecular understanding of this important process.

## 4. Mechanism of quinone reduction

The X-ray structures of complex I lack a Q molecule resolved at the binding site, but central residue interactions have been biochemically identified.[104] In order to construct a Q-bound complex I model, the binding site can be computationally probed by searching for internal protein cavities, followed by relaxation of the Q molecule within the cavity in different redox and protonation states. Such simulations[36,38] identified a binding mode where the Q forms hydrogen-bonding interactions with a protonated His-38 (HisH$^+$) and Tyr-87 (Figure 3A), and where the former is further stabilized by the anionic Asp-139. The Q head group is located *ca.* 20 Å above the membrane plain, and its isoprenoid tail extends all the way to the membrane phase (see Figure 1) in the unusual Q-tunnel, with one side comprising non-polar residues and the other side comprising many charged Glu/Arg ion-pairs.[10,12] In addition to this hydrogen-bonded binding mode, the simulations suggest that Q can also bind in an alternative conformation, where the Q head group forms a stacking interaction with His-38, while retaining its hydrogen bond with Tyr-87 (Figure 3B).[38] DFT calculations further suggest that when Q is oxidized, the stacked conformation is energetically slightly preferred over the hydrogen-bonded conformation. The energetic preference for this stacked conformation becomes somewhat more pronounced upon N2 reduction, suggesting that the redox state of the terminal FeS center might regulate the binding energetics of Q.

QM/MM simulations can be used to probe the eT dynamics between N2 and Q, when both groups and their nearby surroundings are included in the same QM region. Such simulations also require "diabatization" of the initial electron transfer state so that the dynamics is initiated from a state with a reduced electron donor (N2) and an oxidized electron acceptor (Q). In such simulations,[38] the eT between N2 and Q takes place on picoseconds timescales when the latter resides in its relaxed hydrogen-bonded binding mode, whereas no eT was observed in stacked Q-conformation on accessible simulation timescales. Electrostatic PB calculations suggest that the $E_m$ of the Q/SQ redox pair is around -260 mV and -380 mV in the hydrogen-bonded and stacked-binding modes, respectively, indicates that the eT is exergonic in the hydrogen-bonded binding mode, and endergonic in the stacked-binding mode. Importantly, these redox potential calculations support results from earlier electrochemical measurements,[105] which indicate that the $E_m$ of Q/SQ is <-300 mV in the Q-binding site, since no semiquinone radicals was observed. Mechanistically, these findings imply that there is no significant redox-drop between NADH (-320 mV) and Q (-300 mV) when the latter is bound near N2. However, since Q in membranes have an $E_m$ of *ca.* +90 mV, movement of Q towards the membrane would be expected to couple to the release of *ca.* 400 mV (*ca.* 9 kcal mol$^{-1}$) redox-energy, which in turn could thermodynamically drive pumping of two protons across an electrochemical proton gradient of *ca.* 200 mV. The Q tunnel contains several aromatic and charged conserved residues, which could form transient cation-π and/or π-π interactions with Q. If Q indeed would have a second binding-site in this tunnel, it could explain findings from earlier EPR studies, which suggest that SQ resides <15 Å and/or *ca.* 30 Å from the N2 center.[106] These findings might favor a mechanistic model where the piston-like dynamics of Q within its tunnel could couple to the proton pumping process.[4,34]

After formation of SQ, the second electron is transferred along the FeS chain to re-reduce N2, followed by eT between N2 and SQ. In QM/MM simulation of the first eT step, SQ remains deprotonated (Figure 3C). However, the second eT step from N2 to SQ, leads to a stepwise deprotonation of Tyr-87 and His-38, in a process that results in the formation of QH$_2$. Interestingly, electrostatic calculations suggest that N2 reduction is coupled to the protonation of His-169 near the N2 cluster, consistent with earlier experiments on the pH-dependence of the N2 centers, and lack of this pH-dependence in the H169M mutant.[102] Simulations suggest that the second eT step is strongly favored by the deprotonation of His-169, a process that could kinetically control the rate of Q reduction.

## 5. Redox-linked conformational changes in the membrane domain

The DFT-based QM/MM MD simulations on the $QH_2$ formation process, described above, can be used to probe the dynamics on some tens of picosecond timescales, whereas classical MD simulations are necessary for probing the dynamics on longer (ns-μs) timescales. Classical MD simulations show that formation of $QH_2$ by proton transfer from Tyr-87 and His-38 increases the dissociation probability of the Asp-139/His-38 pair. This in turn induces conformational changes in carboxylates and arginines along the E-channel,[10] that propagate to Glu-213/Glu-163, located on a flexible loop in subunit NuoH (Nqo8/ND3).[36] Interestingly, Glu-213 has been refined in a different conformation in complex I from *Y. lipolytica*,[12] supporting that the residue might indeed undergo conformational changes. Electrostatic calculations further suggest that the conformational changes of these glutamates lead to an increase in their protonation probability, which may lead to uptake of protons from the N-side of the membrane. In order to probe the involvement of Asp-139 in triggering this force propagation from the Q-site, the residue was mutated to asparagine *in silico* and *in vitro*. Pumping experiments on the D139N mutant show that the activity of complex I is inhibited by 75% relative to the wild type, supporting the important putative function of this residue.[36] Moreover, QM/MM simulations suggest that two-electron reduction leads to formation of $QH^-$ instead of $QH_2$, since His-38 is likely to be neutral in D139N. Moreover, the simulations show that the resulting $QH^-$ species moves *ca*. 10 Å downwards in the Q-cavity, mimicking in part the conformational changes around the Asp-139 residue, possibly due to repulsion from the anionic Tyr-87 (TyrO$^-$). Although not yet extensively studied in MD simulations, SQ ($Q^{•/-}$) electrostatically resembles such $QH^-$ species and could also trigger similar coupled conformational and electrostatic changes within the E-channel. More experimental data is, however, currently required to determine whether the one- or two-electron reduced Q species are involved in triggering proton pumping.[107-110]

## 6. Function of the proton pump

Classical MD simulations suggest that the membrane domain of complex I undergoes significant hydration changes on *ca*. 200-400 ns timescales.[35,37] These hydration changes are triggered by the protonation states of buried charged residues, to which quasi-one dimensional water chains provide hydrogen-bonded connectivity from the N-side of the membrane. These residues are located in the antiporter-like subunits, and provide also further connectivity along terminal Lys/Glu residues within each subunit to the P-side of the membrane (Figure 4). QM/MM simulations can be further used to probe the proton conduction properties of such classically formed water wires. To this end, the water molecules within the wire and their nearby surroundings are moved into a QM region and the water molecule near the bulk surface modified into a $H_3O^+$ species, while modeling the remaining protein surroundings at the classical force field (MM) level. QM/MM MD simulations of such states further support that the water chains provide effective catalyst for Grotthuss-type proton transfer reactions (Figure 4, *inset*), and also that lateral proton transfer along the antiporter-like subunits is indeed possible.[37]

MD simulations suggest that the proton channels are established at four symmetry-related locations,[37] with an input site near TM7b and output near TM12b in NuoN, NuoM, and NuoL. A "fourth" input channel from the N-side is also observed in the NuoH subunits, forming a hydrated structure near TM6. This hydrated region forms contacts with the glutamate region of the E-channel, which undergo conformational and protonation changes as a result of the Q-reduction process (see above). This region is also close to the Q-tunnel (see above), suggesting that movement of the Q within its tunnel might be strongly coupled to the proton-pumping machinery.

Importantly, the simulations show that the continuous connectivity between the N- and P-sides is broken by bulky phenylalanine and leucine residues along the hydrophilic axis in each antiporter-like, suggesting that complex I strictly regulates the hydrated connectivity across the membrane. These bulky residues might provide important gating points that prevent the protons from leaking to the wrong side of the membrane and help in establishing an efficient pumping machinery. Such residues thus also provide important targets for future site-directed mutagenesis experiments.

Many carrier-type transporters operate by large-scale conformational changes that provide alternate access across the membrane.[111,112] Despite their homology to Mrp-type (multi-resistance and pH-regulated) transporters,[10,113] the antiporter-like subunits in complex I seem to undergo subtle conformational changes in their broken helices and surrounding elements upon the hydration changes.[35,37] Interestingly, the hydration state in complex I can also be modulated by perturbing the structure near these helix elements, suggesting that there is, nevertheless, a connection between these two processes. The channel hydration state is also sensitive to the protonation states of the central lysine residue, as well as to the conformational state of the conserved Glu-Lys ion-pair of each antiporter-like subunit (Figure 1, *inset*). This ion-pair is broken in the X-ray structures of complex I, whereas modeling of the charged state (Glu$^-$/Lys$^+$) results in closure of the ion-pair, with occasional inter-subunit contacts between the NuoN-NuoM and NuoL-NuoM subunits.

Mechanistically it is important to understand how complex I employs direct (electrostatic) and/or indirect (conformational) coupling principles to drive the proton pumping across the membrane.[1-5,34,107-110] Electrostatic calculations suggest that while the interaction between charged residue pairs is very strong, up to 20 kcal mol$^{-1}$, coupling beyond residue pairs may not allow for large enough p$K_a$ modulations that would effectively release the proton across the membrane.[35,37] However, inter-subunit contacts that form between glutamates and lysines in neighboring (NuoN-NuoM and NuoM-NuoL) subunits might help in releasing the proton to the P-side. Due to the relatively weak coupling beyond residue pairs, this may imply that certain pumping models, *e.g.*, the wave-spring model,[5,21] where protons are released in synchronized steps form NuoL/NuoN and NuoM/NuoH, may require indirect (conformational) coupling. Simulations also show that the dissociation of the Lys$^+$/Glu$^-$ ion pair decreases the proton affinity of the central lysine residue, whereas further deprotonation of this residue leads to a decreased water access from the N-side. These findings may provide an important clue into the function of the pumping machinery (see below).

## 7. Putative model for redox-driven proton-pumping

Results from the molecular simulations provide important mechanistic insights that can be used to derive a minimal mechanistic model for the function of the proton pump, schematically shown in Figure 5.[35-38] In this model, reduction of Q triggers conformational changes in NuoH that lead to uptake of a proton from the N-side by water chains, as suggested by the MD simulations.[37] Although the exact details are still unclear, this could in turn increase the dissociation of the Glu/Lys ion-pair in NuoN, which would further lead to proton transfer towards the P-side due to destabilization of the protonated "middle" Lys. Redistribution of the charge in NuoN would in turn dissociate the Glu/Lys ion-pair in NuoM, which would trigger a similar electrostatic perturbation, and lead to an internal proton transfer along the subunit. This charge redistribution would then propagate to NuoL, induce a Glu/Lys dissociation, followed by a proton release from the middle Lys. The MD simulations indicate that deprotonation of the central lysine might close the contact with the N-side, which is further expected to prevent the "released" proton from leaking towards its thermodynamically favored N-side of the membrane. Based on thermodynamic arguments discussed above, movement of the reduced quinone species towards the membrane domain is expected to lead to an energy transduction event, which could push out the proton(s) to the P-side and re-establish the initial state of the catalytic machinery. The sequential propagation of such conformational and electrostatic changes along each antiporter-like subunit could, at least in part, explain why mutations of conserved residues in NuoL inhibit the Q-reduction activity. In this putative model, the conformation of the Lys-Glu ion-pair in TM4-8 is strongly coupled with the proton transfer along the middle Lys → terminal Lys (Glu) in TM9-13, and vice versa. Introduction of residues that disturb either of these processes would therefore be expected to affect the global pumping energetics, as indeed observed in certain NuoL mutants.[7,25,26]

## 8. Conclusions

The respiratory complex I is a redox-driven proton pump that couples a 100 Å electron transfer along its hydrophilic domain to proton transfer across its membrane domain, up to 200 Å away from its active site. Multi-scale molecular simulations can provide important insight into the energetics and dynamics of this remarkable long-range proton-coupled electron transfer (PCET) process. Simulations show that complex I employs coupled conformational and electrostatic changes that trigger $pK_a$ shifts in buried conserved residues. This in turn controls the formation of proton-conducting water wires that provide alternate access between the two sides of the membrane. The pumping machinery in complex I is thermodynamically driven by the redox state and dynamics of the quinone, and it is therefore important to elucidate the exact chemical character of transient intermediates that trigger the proton pump. Molecular simulations provide valuable methods to probe the energetics and dynamics of intermediate states of the catalytic cycle, and the function of central protein residues that are involved in central steps. This information can be used for the design of new biochemical and biophysical experiments. The combination of computational and experimental techniques will be central in elucidating the function of the intricate pumping machinery in complex I.

## 9. Acknowledgement

I acknowledge insightful discussion on complex I with Prof. Mårten Wikström, Dr. Gerhard Hummer, Dr. Vivek Sharma, Dr. Ana P. Gamiz-Hernandez, and Andrea Di Luca. Our work on biological energy conversion machineries has been supported by grants from the German Research Foundation (DFG), the Jane and Aatos Erkko Foundation, the German Academic Exchange Service (DAAD), and the European Research Foundation (ERC). Computing time for our research on complex I is provided by the Leibniz Rechenzentrum/SuperMUC.

**Abbreviations**

B3LYP - Becke's three-parameter functional - a density functionals
DFT - density functional theory
$E_m$ - midpoint redox potential
EOM - equations-of-motion
eT - electron transfer
FeS - iron sulfur center
λ - reorganization energy
MD - molecular dynamics
MM - molecular mechanics
PB - (electrostatic) Poisson-Boltzmann calculations
PCET - proton-coupled electron transfer
$pK_a$ - acid dissociation constant
pT - proton transfer
QM - quantum mechanics
QM/MM - hybrid quantum mechanics/molecular mechanics

**Figures**

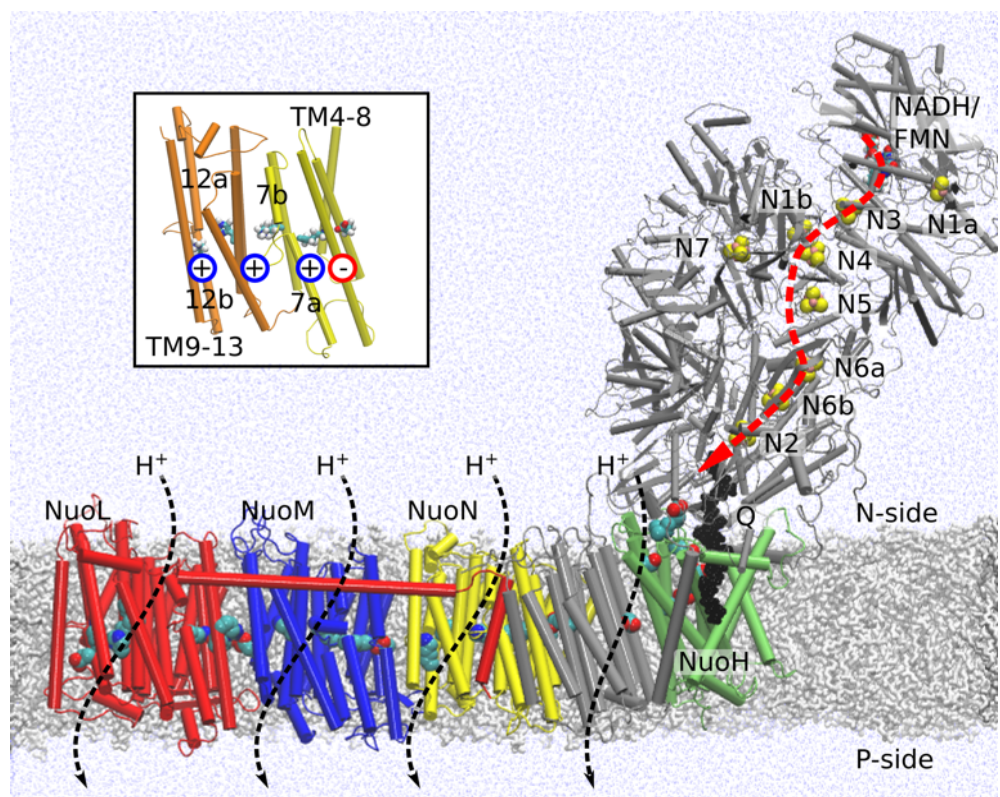

**Figure 1.** A) The structure and function of complex I. The figure shows an MD simulation setup of complex I from *Thermus thermophilus* (PDB ID:4HEA) in a water-membrane-ion environment, comprising *ca*. 840,000 atoms. Q (in black) has been modeled into the structure. Electron transfer from NADH/FMN via the FeS centers (red dotted line) to Q activates the proton pumping in the antiporter-like subunits NuoL (in red), NuoM (in blue), NuoN (in yellow), NuoH (in green). Conserved titratable residues along the membrane domain are also shown in van-der-Waals representation. *Inset*: the structure of TM4-8 and TM9-13 of NuoN. Each antiporter-like subunit (NuoN/M/L) comprises a conserved Lys (indicated with "+")/Glu (indicated with "-") ion-pair, a central Lys (+) and a terminal Lys (+, or Glu in NuoM).

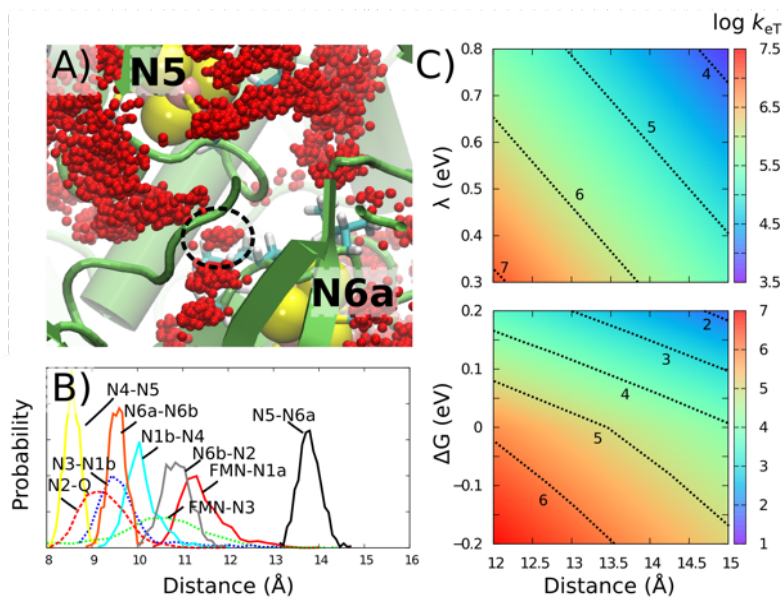

**Figure 2.** A) Averaged dynamics of water molecules (in red) near N5 and N6a suggest that water molecules may bridge empty gaps in the X-ray structure of complex I. B) *Edge-to-edge* distances between the eT centers during a microsecond MD trajectory. C) Sensitivity of the rate-limiting eT rate (log $k_{eT}$) between N5-N6a on the eT parameters, $R$, $\lambda$, and $\Delta G$.

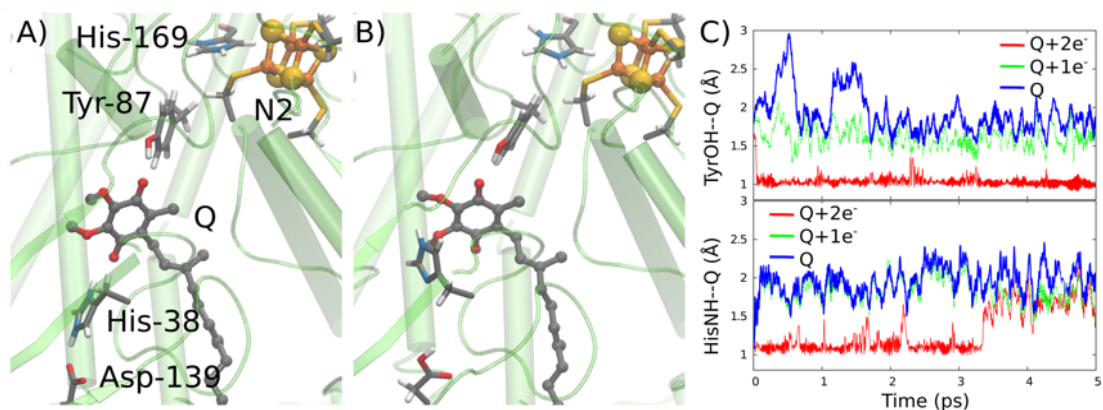

**Figure 3.** Q-binding site in complex I. A) Hydrogen-bonded and B) stacked binding modes of Q, forming contacts with Tyr-87 and His-38. C) QM/MM MD simulations of Q in oxidized state (in blue), one-electron reduced state (Q+1e⁻, in green), and two-electron reduced state (Q+2e⁻, in red). The data suggest that two-electron reduction of Q leads to formation of $QH_2$ by proton transfer from Tyr-87 and His-38. Data in C is obtained from Ref.[35]

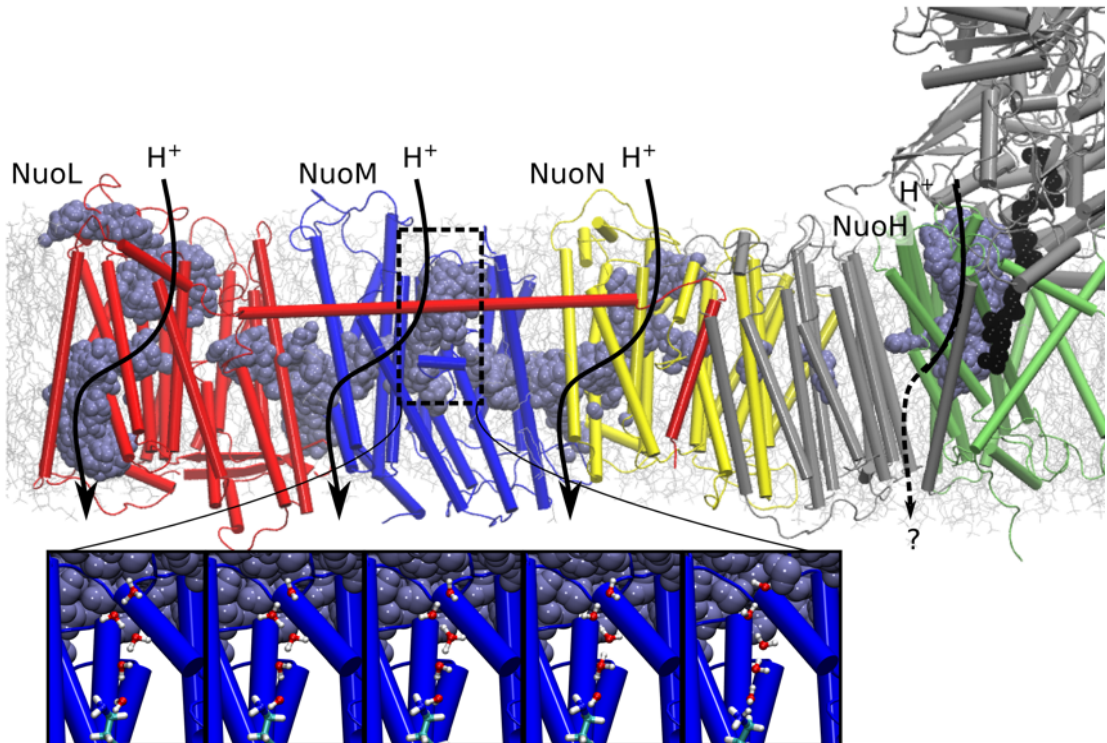

**Figure 4.** Formation of hydrogen-bonded water arrays that establish protonic connectivity across the membrane domain based on microsecond MD simulations of complex I from *Thermus thermophilus*. Proton-channels are formed at four symmetry-related locations in NuoL, NuoM, NuoN along TM7a and TM12b, and along the E-channel region in NuoH (see text). *Inset*: QM/MM MD simulations on Grotthuss-type proton transfer along the classically formed water chains. The proton transfer process takes place on picosecond timescales in water wires formed in the μs-MD trajectories.

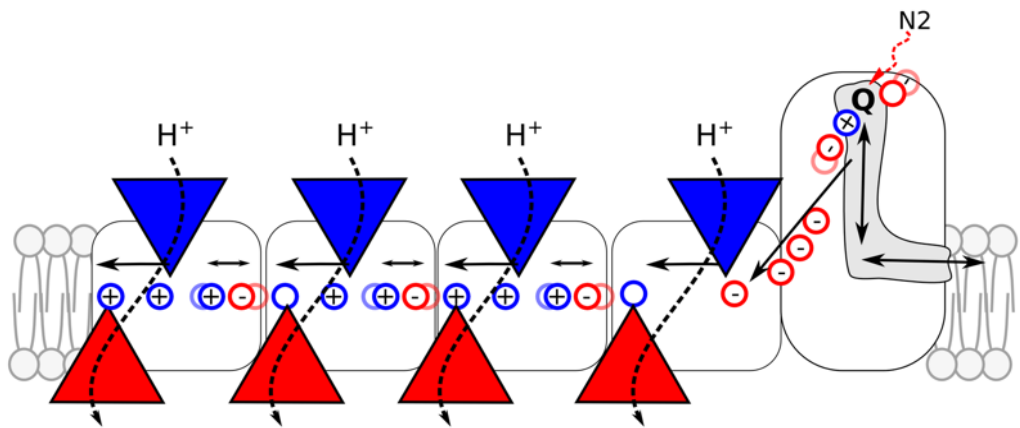

**Figure 5.** Putative schematic pumping model in complex I. Q reduction triggers local electrostatic changes in the active site that propagate to the NuoH subunit, which leads to proton uptake by water wires (blue triangle). Intrinsic proton transfer reactions induce conformational changes in the Lys-Glu ion pair of each antiporter-like subunit, and opens up proton uptake from the N-side (blue triangle) by sequential propagation along the membrane domain. Movement of Q along its tunnel (in gray) couples to release of redox energy that is employed to push the proton(s) across the membrane.